\documentstyle[12pt]{article}
\def\be{\begin{equation}}
\def\ee{\end{equation}}

\begin{document}

\baselineskip 4ex

\centerline{QRPA plus Phonon Coupling Model and the Photoabsorbtion
            Cross Section for $^{18,20,22}$O~}

\vspace*{0.8cm}

\centerline{G. Col\`o\footnote{
                             Tel.: + 39 - 02 - 58357261; 
                             fax: + 39 - 02 - 58357487; 
                             e-mail: colo@mi.infn.it} 
        and P.F. Bortignon}

\centerline{\it Dipartimento di Fisica, Universit\`a degli Studi, }  
\centerline{\it and INFN, Via Celoria 16, I-20133 Milano (Italy) }

\vspace*{0.8cm}
\noindent

\centerline{ABSTRACT}
We have calculated the electric dipole strength distributions in the
unstable neutron rich oxygen isotopes $^{18,20,22}$O, in a model which
include up to four quasi-particle-type configurations. The model is the
extension, to include the effect of the pairing correlations, of a previous
model very successful around closed shell nuclei, and it is based on the
quasi-particle-phonon coupling. Low-lying dipole strength is found,
which exhausts between 5 and 10\% of the Thomas-Reiche-Kuhn (TRK) 
energy-weighted-sum-rule (EWSR) below 15 MeV excitation energy, in rather good
agreement with recent experimental data. The role of the phonon coupling 
is shown to be crucial in order to obtain this result.

\medskip
\centerline{PACS numbers: 24.30.Cz, 21.60.Jz. } 
\centerline{Keywords: low-lying dipole strength, giant resonances, }
\centerline{neutron-rich nuclei, quasi-particle RPA, phonon coupling. }

\newpage

\section{Introduction}
\label{intro}

The study of Giant Resonances (GR) in nuclei far from the stability line, such as halo and
skin nuclei~\cite{Tan,Pgh}, is still in its infancy.
The observation of large E1 strength at low excitation energy, just above threshold in light
drip-line nuclei, has produced strong excitement and large amount of 
activities, both experimental and 
theoretical. Electromagnetic dissociation measurements of this low-lying stength have been 
performed for the neutron halo nuclei $^{6}$He~\cite{He6}, $^{8}$He~\cite{He8},$^{11}$Li~\cite{Li}, 
$^{11}$Be~\cite{Be11}, $^{12}$Be\cite{Be12}, $^{14}$Be\cite{Be14}, 
$^{19}$C~\cite{C}, and for the 
proton halo nuclei
$^{8}$B~\cite{B} and $^{13}$O~\cite{O}. From the theory side, a quantum-mechanical threshold 
effect was shown to enhance the transition strength from the loosely bound nucleons 
to the continuum ~\cite{Pgh1,Ca}, essentially produced by the optimal matching of the 
wavelength of the scattering states 
with the large penetration length into the forbidden region of the 
weakly bound orbitals. In the case of two-nucleon halo  nuclei, the enhancement due
to the coherence in the transition amplitudes between the loosely bound nucleons, and also 
between these and the core, has been found important in recent shell-model calculations with 
the use of extended wave functions for the matrix elements of the dipole 
operator~\cite{Suz,Sa1} (see also~\cite{Esb92}).

Moving to slightly heavier, 
eventually skin nuclei, the first attempt~\cite{Aum} of a systematic 
experimental study of GR has been recently performed at GSI,  with the investigation of
the dipole response of unstable neutron rich oxygen isotopes up to $^{22}$O and to an 
excitation energy of 30 MeV. Use has been done of the electromagnetic excitation in heavy-ion 
collisions at around 600 MeV/u beam energy. Low-energy collective modes in 
$^{20}$O were also measured recently for the first time by means of inelastic
proton scattering~\cite{Kha}.

Mean time, theoretical studies have been performed in the context of {\it mean-field} theories, 
like self-consistent Hartree-Fock and Random Phase Approximation (HF+RPA) 
calculations~\cite{Ca23}, eventually 
including exactly the crucial coupling to the continuum~\cite{Hasa,Sa2}, 
or HF-BCS plus quasi-particle RPA (QRPA) calculations~\cite{Kha1} (which 
include the contribution of the pairing force), as well as relativistic
RPA (RRPA) calculations~\cite{Dar},
in order to investigate how the strength distributions of different 
multipole operators, in particular monopole, dipole and quadrupole, are evolving as one
approaches the drip lines. The main features result to be: (a) the unusual 
concentration of 
multipole strength at the continuum 
threshold, already mentioned, when the threshold energy becomes of the order of one or
few MeV, 
and (b) the strong interplay or mixing between the so-called 
isoscalar and isovector modes familiar from the response of stable nuclei. 

Much less, and not at all systematically, has been done {\it beyond mean field},
that is,  
including the coupling of single-particle states to vibrational modes, in particular low-lying 
vibrations~\cite{BBB}, as widely done in the case of stable nuclei in the last decades. 
Some pioneering works on the effects of this coupling on the single-particle self-energy
and the GR strength function
in netron-rich nuclei are reported, e.g., in
Refs.~\cite{Vin,Ghi,Co1,Co2,Lit}. Of course, these correlations beyond mean field are also 
included, although somewhat hidden, in the shell-model calculations 
performed by several groups~\cite{Shm}. 
In particular, we mention     
the calculation of Ref.~\cite{Sa3}, where the dipole strength functions in 
the O isotopes
were calculated, before the results of the GSI experiment quoted above were known.

Thus, it appears timely to extend in the present paper the microscopic model described in
detail in Ref.~\cite{Co3} and aimed to a microscopic description of the excitation 
and decay of GR in stable nuclei, to the case of the neutron rich systems. This is done
in the present work, were pairing correlations are included in the model in a simple
way. The model is based on HF-RPA with Skyrme effective forces and includes on top 
escape and spreading effects in a consistent way. It was found to give accurate 
predictions, around closed shell nuclei, for the excitation energy and strength 
of GR as well
as for widths and particle decay branching ratios~\cite{Pas}. 

The paper is organized as follows. In Section 2, the model and its extension are presented.
The application to the calculation of the dipole strength functions in the O isotopes
is described in Sect. 3, including also the comparison with the recent experimental data
of Ref.~\cite{Aum} and with the shell-model results of Ref.~\cite{Sa3}. 
Preliminary results were published in conference proceedings~\cite{Co4}.
Summary and 
conclusions 
are drawn in Sect. 4.
 
\section{Description of the model}
\label{model}

In this section, we shortly present the extension of the model of Ref.~\cite{Co3},
to include in a minimal way the effects of the pairing interaction. Indeed, we are 
interested to perform calculations along isotopes chains and the inclusion of pairing 
correlations is called for.   

The starting input is an effective Hamiltonian $H$ which includes a two-body interaction 
of the Skyrme type~\cite{Sky}. The self-consistent mean field is determined by 
solving the 
Hartree-Fock (HF) equations on a radial mesh, coupled with the standard 
BCS equations for the added pairing interaction (within the constant pairing 
gap approximation). This method is well known for spherical 
nuclei in the case of Skyrme interactions~\cite{Sky,Vau}. 
The unoccupied states, at negative as well as positive
energy, are determined by using a harmonic oscillator basis, so the continuum
is discretized. 
This procedure defines a subspace labelled by $Q_1$ and including discrete 
two quasi-particle (2qp) configurations. The standard~\cite{Row} QRPA equation
in matrix form are solved on a basis of $J^\pi$=1$^-$ configurations belonging
to $Q_1$, to obtain collective (or non-collective) low-lying and GR modes
(see also Ref.~\cite{Kha1}).
Special attention is paid to project out of the spectrum the 
spurious center-or-mass state, by renormalizing by a few percent the matrix 
elements of the residual interaction in such a way that the spurious state
is pushed to zero energy. 

To account for the spreading width $\Gamma^{\downarrow}$ we build a second 
subspace $Q_2$ (the subspace $P$ introduced in Ref.~\cite{Co3} to account 
for the escape width $\Gamma^{\uparrow}$ is not included in the present 
model, see also next section). In $Q_2$, there are the main configurations 
known~\cite{BBB} to play a major role in the damping process of GR: they 
are 2qp states coupled to a collective vibration calculated using the 
discrete QRPA in the $Q_1$ subspace as described above. We discuss in the
next section what vibrations have been selected and included in $Q_2$. 

Using the projection operator formalism, it is easily found that 
the effects of  
coupling the subspace $Q_2$ to $Q_1$ are described by the following effective 
Hamiltonian ${\cal H}$ acting in the $Q_1$ space
\begin{eqnarray}
     {\cal H} (E) & \equiv & Q_1 H Q_1 
     + W^\downarrow(E) \nonumber \\
     \ & = & Q_1 H Q_1 + 
     Q_1 H Q_2 {\textstyle 1 \over \textstyle
     E - Q_2 H Q_2 + i\epsilon}
     Q_2 H Q_1, \nonumber \\
     \ & \ &
\label{H_eff}\end{eqnarray}
where $E$ is the excitation energy. For each value of $E$, the QRPA equations 
corresponding
to this complex Hamiltonian ${\cal H}(E)$ are solved (this is done using
the basis of the discrete QRPA states, eliminating those with negligible
strength). The resulting sets of complex eigenstates
$|\nu\rangle$ and eigenvalues $E_\nu-i\Gamma_\nu/2$, 
enable to calculate all relevant quantities, in particular 
the strength function $S(E)$
of the operator $\hat F$ of interest:
\be
S(E)=-{1\over \pi}\ Im\ \sum_\nu {\langle g.s. | \hat F | \nu \rangle^2\over
{E-E_\nu+i\Gamma_\nu/2}}.
\ee 

More in detail, each matrix element of $W^{\downarrow}(E)$ written in the basis 
of the discrete QRPA states is a linear combination of matrix elements 
$W^{\downarrow}_{\alpha\beta,\gamma\delta}(E)$ in the basis of the 2qp configurations
$(\alpha\beta)$, $(\gamma\delta)$ etc. These latter matrix elements are sums of the
eight terms depicted in Fig.~1. 
To evaluate these diagrams, we employ the following expression for the particle-vibration 
coupling Hamiltonian,
\be
 V = \sum_{\alpha\beta} \ \sum_{LnM} \langle \alpha |
 \varrho^{(L)}_n (r) v(r) Y_{LM}(\hat r) | \beta \rangle
 \ a^\dagger_\alpha a_\beta,
\label{pvc}\ee
and we use the standard BCS expressions to relate the quasi-particle operators
$c^\dagger_\alpha$ and $c_\alpha$ to the single-particle ones 
$a^\dagger_\alpha$ and $a_\alpha$ (see Eq.~(\ref{BCS})).
In Eq.~(\ref{pvc}) the vibration (phonon) $|n\rangle$ is characterized by its angular
momentum $L$ and its transition density $\varrho_n^{(L)}(r)$, while the form factor 
$v(r)$ is related to the particle-hole interaction $V_{ph}$ derived from
the Skyrme force as $V_{ph}(\vec r_1,\vec r_2)=v(r_1)\delta (\vec r_1,\vec r_2)$.
The detailed expressions of the diagrams are given in the Appendix.

\section{Application to the dipole response of $^{18-22}$O}
\label{results}

We use the model discussed above to calculate the electric dipole strength 
distributions in the unstable neutron rich oxygen isotopes. 
As anticipated in the introduction, in a recent experiment~\cite{Aum} at GSI its 
knowledge has been extended to the isotopes heavier than $^{18}$O, for which it was 
well known~\cite{Woo,Kne}, with systematic measurements from $A$=17 to $A$=22. 
Low-energy strength was detected,
which exhausts up to 12\% of the classical Thomas-Reiche-Kuhn (TRK) energy-weighted 
sum rule (EWSR) at excitation energies below
15 MeV (approximate threshold for the emission of the protons, which are not detected 
in the experiment). The comparison 
with these data constitutes an adequate test for our model.  

The calculations have been performed in the isotopes
$^{18,20,22}$O, and using the effective force SIII~\cite{Sky}. 
This force, in the HF approximation, reproduces very well the binding energy of these 
nuclei (although it fails to fix the drip line for $Z$=8 at $A$=24). However, 
the corresponding self-consistent RPA description of the low-lying vibrations 
is rather 
poor in the comparison with the experimental data, and is 
improved by the inclusion of some pairing correlations~\cite{Kha,Kha1}. 
Therefore, we do 
include a pairing force with constant matrix elements of strength $G=0.4$ MeV 
(about 30\% of the standard value), and adopt the corresponding values of 
$\Delta_n$ for the neutrons 
in our constant gap calculation. These pairing gaps are 
smaller than 1 MeV (0.74, 0.79 and 0.21 MeV respectively in $^{18,20,22}$O), and the binding 
energies differ only by about 1\% from the HF value. The $B(E2)$ values for the
low-lying 2$^+$ states are somewhat improved compared to the RPA calculation 
(in the case 
of $^{20}$O, the experimental value for the $B(E2)$ is 
28$\pm$2 e$^2$fm$^4$~\cite{Ram}, RPA gives only about 3 e$^2$fm$^4$ and 
our QRPA with the small gap gives 15.6 e$^2$fm$^4$).

The HF-BCS calculations are done in coordinate space, using a radial grid
of 0.1 fm extending up to 15 fm. The use of a constant pairing gap $\Delta$
leads to unrealistic results unless a cutoff is introduced in the single-particle
space, in such a way that states above this cutoff do not feel any pairing
interaction: in our case, the cutoff is just above the 1d$_{3/2}$ neutron state.
In the QRPA calculations for the dipole and for the isoscalar modes to be
included in the subspace $Q_2$, the residual interaction between
quasi-particles is derived from the Skyrme force without including the
pairing contribution. The 2qp basis is chosen large enough so that
the appropriate EWSR are satisfied. In the case of the dipole, we refer to
the double commutator sum rule which is enhanced with respect to the classical
TRK sum rule by a factor 1.33 in the case of the SIII interaction at hand. 

In the QRPA calculation, the dipole strength below 15 MeV is found smaller
than the experimental one (see 
Table~\ref{table1}). Then, the full 
coupling to the complex configurations of the $Q_2$ space is taken into 
account. The phonons included are those of multipolarity 1$^-$, 2$^+$, 3$^-$ 
and 4$^+$ which lie below 30 MeV and absorb more than 5\% of the total 
strength. (In the case of $^{18}$O, the 0$^+$ modes have been included but
their effect is found to be rather small). The final results for the 
integrated cross section up to 15 MeV is shown in Table~\ref{table1} 
in comparison 
with the experimental data~\cite{Aum} and with the results of the shell-model 
calculation of~\cite{Sa3}. 
The coupling with phonons increases 
the low-energy cross section, bringing it in rather good agreement with the 
data. Moreover, the drop of the cross section in $^{22}$O is also reproduced.
In the case of $^{18}$O our results are also remarkably similar to the 
shell-model ones, suggesting that we are including the most important 
correlations.  

The entire cross sections obtained with the full coupling for the three 
isotopes are displayed in Fig.~2. 
The progressive cumulation of cross section at lower energies 
with increasing A
is evident, as well as the mild dependence on A of the main GDR peak and the
appearance of more fragmented and peculiar structures at higher energies.
In the case of $^{18}$O, the peak energy at 24.2 MeV compares very well with 
the known experimental value~\cite{Woo}, although the full experimental 
strength distribution looks wider (see Fig.~3(e) of~\cite{Woo}).   

These features are not qualitatively very different in the 
calculations we performed with much larger pairing gap values, that is, 
the standard $12/\sqrt {\rm A}$ MeV values. 
The results of these latter calculations are displayed in Fig.~3.
In this case, compared to the calculation with the small gaps, 
the GDR main peak is more damped and the high energy structures more
pronounced. The largest difference appears for the case of $^{22}$O: 
moreover, in this nucleus
the strength below 15 MeV is definitely larger than in the small gap case, at
variance with $^{18,20}$O where this strength is not very sensitive to the
value of $\Delta_n$ we have employed (see Table~\ref{table1}). 
The fact that the GDR peak energy is not markedly affected by the size
of the pairing gap in all cases can be understood since this is essentially ruled by
mean field effects and at that level, 
the effect of pairing should not be 
important for states at energies much larger than $\approx 2\Delta$. Concerning
the coupling with the phonons, for large values of the pairing gap, the low-energy 
surface vibrations become more collective but tend also 
to be pushed at higher energies (see also Fig.~1 of Ref.~\cite{Kha1}) and these two
facts have opposite consequences on the effectiveness of the phonon 
coupling. In the present case we understand from Fig.~3 that 
the first of these two effects (increased phonon collectivity) seems to
be somewhat dominant. 

We have not discussed so far the sensitivity of the results to the choice
of the effective Skyrme force. An example can be seen in Fig.~\ref{forces}, 
where the 
full and dashed lines are two calculations for $^{18}$O with the large gap, 
using respectively the forces SIII and SGII~\cite{SGII}. The difference between the 
two curves is essentially accounted for by a downward shift of about 2 MeV.
This is mainly due to the different effective mass associated with the two
interactions. 

The most important outcome of our study is that
the effect of the coupling with complex configurations including one phonon is
very large and unavoidable if the comparison with experimental data is envisaged.
The point is clearly emphasized in Fig.~5, for $^{18}$O. 
The mean field (i.e., QRPA) 
result, with the simple smearing by means of 1 MeV width Lorentzian functions,
are qualitatively altered when the coupling with phonons is included. This
leads to a largely spread cross section, in much better agreement with the experiment.
The effect is certainly larger than in stable, heavy nuclei, where most of our experience
is. One main reason is the lack of the large cancellation~\cite{BBB}
between the diagrams in the upper and lower two lines of Fig.~1. 
This is due to the very asymmetric phase-space available, 
in these light nuclei, for
the hole-like and particle-like contributions to $W^\downarrow$ and was already discussed in Ref.~\cite{Ari}
for the shell-model calculation of the giant quadrupole resonance (GQR) in $^{16}$O compared
to $^{40}$Ca. This asymmetry goes together, in hindering large 
cancellations, with the small number of components 
in the wave functions of the low-lying structures, as reported in Table~\ref{table2}
and in Table~\ref{table3}. Consequently, these structures 
should {\em not} be considered
of collective nature.  

Admittedly, this small number of components makes more critical than in the heavier nuclei
the omission of the coupling to the continuum. The neutron separation energy in
$^{22}$O is still large, of the order of 7 MeV, but certainly the large role played
in the wave functions by neutron quasi-particles like the 2p$_{3/2}$ in the continuum   
poses some questions, worthwhile to be investigated. In any case, we show in Fig.~6 
the results at the level of HF (that is, the unperturbed dipole strength function) 
obtained either 
taking properly the continuum into account or discretizing it. By choosing
reasonable parameters to average the discrete results (see the caption to the figure)
we reproduce very well the exact results, and this means that peak regions and 
corresponding intensities can be, in this case, accounted for by the discrete
calculation.  

\section{Conclusion}
\label{conclu}

In the present work we have calculated the electric dipole strength 
distributions in the
unstable neutron rich oxygen isotopes $^{18,20,22}$O. The model we used
is the extension, to include the effect of the pairing correlations, 
of a previous model~\cite{Co3} which has been very successful 
around closed shell nuclei, and which is based on the
quasi-particle-phonon coupling. Thus, the 2qp configurations 
of the standard QRPA approach, are coupled to 4qp-type  
configurations built with two uncorrelated quasi-particles plus a
collective phonon (these are known to be 
particularly efficient in redistributing the GR strength).

Low-lying dipole strength is found,
exhausting between 5 and 10\% of the TRK sum rule
below 15 MeV excitation energy, in rather good
agreement with the recent experimental data of~\cite{Aum}.   
In order to obtain this result, and more generally to spread out the
whole strength distribution including the main GDR peak,  
the role of the coupling with phonons appears to be crucial: 
therefore, the importance of particle-vibration coupling in neutron-rich
nuclei is one of the most important outcomes of our paper. 
The comparison with the strength distributions obtained
by means of large scale shell-model calculations~\cite{Sa3} 
is satisfactory in terms of integrated quantities, especially in $^{18}$O, 
although the detail of the distributions looks quite different.

Confident in our model, we are planning to apply it in the case of 
much heavier nuclei, relevant, e.g., for the study of the r-process and 
nucleosyntesis. In these nuclei, other types of microscopic calculations
like the full shell-model may result prohibitive for computational reasons and,  
to describe the low-lying dipole strength, macroscopic 
models~\cite{Iked,VanI} which may 
not contain all proper physical ingredients have been used so far~\cite{Gor}. 

\section*{Acknowledgments}

We especially thank 
T. Aumann, H. Emling and A. Leistenschneider, for extensive communication and
explanations about the detection of the low-lying dipole strength in the
oxygen isotopes. We also thank M. Thoenessen and Nguyen Van Giai for
stimulating discussions. P.F.B. acknowledges RIKEN for financial support and
nice hospitality during the period in which part of this work was
performed, and the enlightening discussions about it with K. Ikeda.

\section*{Appendix}

In this Appendix we provide the expressions of the eight 
diagrams contributing to the 
matrix element $W^{\downarrow}_{\alpha\beta,\gamma\delta}(E)$ mentioned in Sec. 2 and depicted in Fig.~1.

\begin{eqnarray}
W^{\downarrow}_{I} & = & \delta(\alpha,\gamma)\delta(j_\beta,j_\delta)
[1+\delta(\alpha,\beta)]^{-1/2}[1+\delta(\alpha,\delta)]^{-1/2}(2j_\beta+1)^{-1}
\sum_{\sigma,Ln} 
[1+\delta(\alpha,\sigma)]^{-1/2}
\hfill\nonumber \\
&& 
(U_\beta U_\sigma+V_\beta V_\sigma)(U_\sigma U_\delta+V_\sigma V_\delta) 
\int dr u_\beta(r)u_\sigma(r)\varrho_n^{(L)}(r)v(r)
\int dr u_\delta(r)u_\sigma(r)\varrho_n^{(L)}(r)v(r) \hfill\nonumber \\
&&
\vert \langle \beta \vert\vert Y_L \vert\vert \sigma \rangle \vert^2\cdot 
[\omega-(E_\alpha+E_\sigma+\omega_n)+i\eta]^{-1}; 
\end{eqnarray}

\begin{eqnarray}
W^{\downarrow}_{II} & = & \delta(\alpha,\delta)\delta(j_\beta,j_\gamma)
[1+\delta(\alpha,\beta)]^{-1/2}[1+\delta(\alpha,\gamma)]^{-1/2}(2j_\beta+1)^{-1}
\sum_{\sigma,Ln} 
[1+\delta(\alpha,\sigma)]^{-1/2}
\hfill\nonumber \\
&& 
(U_\beta U_\sigma+V_\beta V_\sigma)(U_\sigma U_\gamma+V_\sigma V_\gamma) 
\int dr u_\beta(r)u_\sigma(r)\varrho_n^{(L)}(r)v(r)
\int dr u_\gamma(r)u_\sigma(r)\varrho_n^{(L)}(r)v(r) \hfill\nonumber \\
&&
(-1)^{j_\alpha-j_\beta+J}
\vert \langle \beta \vert\vert Y_L \vert\vert \sigma \rangle \vert^2\cdot 
[\omega-(E_\alpha+E_\sigma+\omega_n)+i\eta]^{-1}; 
\end{eqnarray}

\begin{eqnarray}
W^{\downarrow}_{III} & = & \delta(\beta,\delta)\delta(j_\alpha,j_\gamma)
[1+\delta(\alpha,\beta)]^{-1/2}[1+\delta(\beta,\gamma)]^{-1/2}(2j_\alpha+1)^{-1}
\sum_{\sigma,Ln} 
[1+\delta(\beta,\sigma)]^{-1/2}
\hfill\nonumber \\
&& 
(U_\alpha U_\sigma+V_\alpha V_\sigma)(U_\sigma U_\gamma+V_\sigma V_\gamma) 
\int dr u_\alpha(r)u_\sigma(r)\varrho_n^{(L)}(r)v(r)
\int dr u_\gamma(r)u_\sigma(r)\varrho_n^{(L)}(r)v(r) \hfill\nonumber \\
&&
\vert \langle \alpha \vert\vert Y_L \vert\vert \sigma \rangle \vert^2\cdot 
[\omega-(E_\beta+E_\sigma+\omega_n)+i\eta]^{-1}; 
\end{eqnarray}

\begin{eqnarray}
W^{\downarrow}_{IV} & = & \delta(\beta,\gamma)\delta(j_\alpha,j_\delta)
[1+\delta(\alpha,\beta)]^{-1/2}[1+\delta(\beta,\delta)]^{-1/2}(2j_\alpha+1)^{-1}
\sum_{\sigma,Ln} 
[1+\delta(\beta,\sigma)]^{-1/2}
\hfill\nonumber \\
&& 
(U_\alpha U_\sigma+V_\alpha V_\sigma)(U_\sigma U_\delta+V_\sigma V_\delta) 
\int dr u_\alpha(r)u_\sigma(r)\varrho_n^{(L)}(r)v(r)
\int dr u_\delta(r)u_\sigma(r)\varrho_n^{(L)}(r)v(r) \hfill\nonumber \\
&&
(-1)^{j_\alpha-j_\beta+J}
\vert \langle \alpha \vert\vert Y_L \vert\vert \sigma \rangle \vert^2\cdot 
[\omega-(E_\beta+E_\sigma+\omega_n)+i\eta]^{-1}; 
\end{eqnarray}

\begin{eqnarray}
W^{\downarrow}_{V} & = & [1+\delta(\alpha,\beta)]^{-1/2}[1+\delta(\gamma,\delta)]^{-1/2}[1+\delta(\alpha,\delta)]^{-1/2}
(U_\alpha U_\gamma-V_\alpha V_\gamma) (U_\beta U_\delta-V_\beta V_\delta)
\hfill\nonumber \\
&&
\sum_{Ln}   \left\{ \begin{array}{ccc} j_\alpha & j_\beta & J \\
                                       j_\delta & j_\gamma & L
\end{array} \right\} 
\int dr u_\alpha(r)u_\gamma(r)\varrho_n^{(L)}(r)v(r)
\int dr u_\beta(r)u_\delta(r)\varrho_n^{(L)}(r)v(r) \hfill\nonumber \\
&&
(-1)^{j_\beta+j_\gamma+J} 
\langle \alpha \vert\vert Y_L \vert\vert \gamma \rangle \cdot 
\langle \beta \vert\vert Y_L \vert\vert \delta \rangle \cdot 
[\omega-(E_\alpha+E_\delta+\omega_n)+i\eta]^{-1}; 
\end{eqnarray}

\begin{eqnarray}
W^{\downarrow}_{VI} & = & [1+\delta(\alpha,\beta)]^{-1}[1+\delta(\gamma,\delta)]^{-1/2}
(U_\beta U_\gamma-V_\beta V_\gamma) (U_\alpha U_\delta-V_\alpha V_\delta)
\hfill\nonumber \\
&&
\sum_{Ln}   \left\{ \begin{array}{ccc} j_\alpha & j_\beta & J \\
                                       j_\gamma & j_\delta & L
\end{array} \right\} 
\int dr u_\beta(r)u_\gamma(r)\varrho_n^{(L)}(r)v(r)
\int dr u_\alpha(r)u_\delta(r)\varrho_n^{(L)}(r)v(r) \hfill\nonumber \\
&&
(-1)^{j_\beta+j_\gamma} 
\langle \alpha \vert\vert Y_L \vert\vert \delta \rangle \cdot 
\langle \beta \vert\vert Y_L \vert\vert \gamma \rangle \cdot 
[\omega-(E_\alpha+E_\beta+\omega_n)+i\eta]^{-1}; 
\end{eqnarray}

\begin{eqnarray}
W^{\downarrow}_{VII} & = & [1+\delta(\alpha,\beta)]^{-1/2}[1+\delta(\gamma,\delta)]^{-1/2}[1+\delta(\beta,\gamma)]^{-1/2}
(U_\alpha U_\gamma-V_\alpha V_\gamma) (U_\beta U_\delta-V_\beta V_\delta)
\hfill\nonumber \\
&&
\sum_{Ln}   \left\{ \begin{array}{ccc} j_\alpha & j_\beta & J \\
                                       j_\delta & j_\gamma & L
\end{array} \right\} 
\int dr u_\alpha(r)u_\gamma(r)\varrho_n^{(L)}(r)v(r)
\int dr u_\beta(r)u_\delta(r)\varrho_n^{(L)}(r)v(r) \hfill\nonumber \\
&&
(-1)^{j_\beta+j_\gamma+J} 
\langle \alpha \vert\vert Y_L \vert\vert \gamma \rangle \cdot 
\langle \beta \vert\vert Y_L \vert\vert \delta \rangle \cdot 
[\omega-(E_\beta+E_\gamma+\omega_n)+i\eta]^{-1}; 
\end{eqnarray}

\begin{eqnarray}
W^{\downarrow}_{VIII} & = & [1+\delta(\alpha,\beta)]^{-1/2}[1+\delta(\gamma,\delta)]^{-1}
(U_\beta U_\gamma-V_\beta V_\gamma) (U_\alpha U_\delta-V_\alpha V_\delta)
\hfill\nonumber \\
&&
\sum_{Ln}   \left\{ \begin{array}{ccc} j_\alpha & j_\beta & J \\
                                       j_\gamma & j_\delta & L
\end{array} \right\} 
\int dr u_\beta(r)u_\gamma(r)\varrho_n^{(L)}(r)v(r)
\int dr u_\alpha(r)u_\delta(r)\varrho_n^{(L)}(r)v(r) \hfill\nonumber \\
&&
(-1)^{j_\beta+j_\gamma} 
\langle \alpha \vert\vert Y_L \vert\vert \delta \rangle \cdot 
\langle \beta \vert\vert Y_L \vert\vert \gamma \rangle \cdot 
[\omega-(E_\beta+E_\delta+\omega_n)+i\eta]^{-1}. 
\end{eqnarray}

In the above expressions, $\sigma$ labels an intermediate
quasi-particle state, $E$ and $u(r)$ indicate respectively the quasi-particle
energy and the single particle radial wave function whereas $U$ and $V$
are the usual BCS coefficients which relate single particle
and quasi-particle operators according to
\be
c^\dagger_\alpha = U_\alpha a^\dagger_\alpha + V_\alpha a_{\bar\alpha},
\label{BCS}
\ee 
where $\bar\alpha$ is the time-reversed of $\alpha$.
Concerning the vibrational states (phonons),
these are labelled by the multipolarity $L$ and the index $n$: 
$\omega_n$ and $\delta\varrho_n^{(L)}(r)$ 
are the corresponding energies and radial 
transition densities as already discussed in the text in Sec.~\ref{model} 
($v(r)$ is also discussed there). 
The averaging parameter $\eta$ is set at 0.5 MeV in the present calculation. 

The above expressions are derived writing the single
particle wave functions as $(u_{nl}(r)/r)i^l Y_{lm}(\hat r)$
(phase convention II of
Ref.~\cite{Row}). In the limit of zero pairing gap, the
expressions reported in Ref.~\cite{Co3} are recovered, where however,
phase convention I of Ref.~\cite{Row} without the factor $i^l$, was used.

\newpage

\begin{figure}
Fig. 1. Diagrams corresponding to the coupling of the 2qp configurations 
to the more complicated states including a phonon. See Sec.~2 for the
discussion of this coupling, the detailed analytic expressions 
being given in the
Appendix. 
\label{graphs}
\end{figure}

\begin{figure}
Fig. 2. Total photoabsorbtion cross section for the isotopic chain 
$^{18,20,22}$O obtained within the full phonon coupling (using 
small pairing gaps
as discussed in Sec. 3).
\label{full}
\end{figure}

\begin{figure}
Fig. 3. Same as Fig.~2 in the case of large pairing gaps (i.e.,
$\Delta_n$=$12/\sqrt {\rm A}$ MeV). 
\label{largegap}
\end{figure}

\begin{figure}
Fig. 4. Total photoabsorbtion cross section in the case of $^{18}$O and of
the large pairing gap, using two different effective forces namely SIII and
SGII. 
\label{forces}
\end{figure}

\begin{figure}
Fig. 5. Total photoabsorbtion cross section for $^{18}$O. 
The large spreading induced by the coupling with
the phonons included in the complete calculation (full line) contrasts with
the sharp peaks of the QRPA result (dashed line). The integrals of the two
curves over the entire range are respectively 303 and 320 MeV$\cdot$mb. 
\label{meanfield}
\end{figure}

\begin{figure}
Fig. 6. Unperturbed (i.e., Hartree-Fock) dipole strength obtained for
$^{22}$O either using the proper continuum (full line) or 
averaging (dashed line)
the sharp peaks (bars) of the discretized continuum. For the
averaging procedure, Lorentzian functions having width 2.2 MeV and
1.6 MeV have been employed, respectively for peaks below or above 12.5 
MeV. Units are fm$^2$ for the bars and MeV$^{-1}\cdot$fm$^2$ for the lines.
\label{cont}
\end{figure}

\newpage

\begin{table}[t]
\newlength{\digitwidth} \settowidth{\digitwidth}{\rm 0}
\catcode`?=\active \def?{\kern\digitwidth}
\caption{Total photoabsorbtion cross section integrated up to 15 MeV obtained
in our QRPA (second row and third row) and QRPA plus phonon coupling 
(third and fourth row) 
calculations. The different choices for the pairing gap $\Delta_n$ (either small 
or large) are discussed extensively in the text. The results are
compared with the photoneutron cross section measured at GSI (first row) and 
with the shell-model calculation of Ref.~{\protect \cite{Sa3}} 
(last row). All numbers
are in MeV$\cdot$mb. } 
\vspace{0.5cm}
\label{table1}
\begin{tabular*}{\textwidth}{|l|ccc|}
\cline{1-4}
                       & $^{18}$O      & $^{20}$O       & $^{22}$O         \\ 
\cline{1-4}
Experiment             & 22.14 $\pm$ 2.4 & 33.7 $\pm$ 3.5 & 23.22 $\pm$ 2.75 \\
QRPA (small gap)       & 11.43         & 19.28          & 15.07            \\
QRPA (large gap)       & 11.00         & 18.21          & 22.36            \\
QRPA plus phonon coupling (small gap)   
                       & 18.26         & 25.65          & 17.05            \\
QRPA plus phonon coupling (large gap)   
                       & 18.55         & 25.92          & 22.87            \\
Shell model~\cite{Sa3} & 17.14         & 31.26          & 30.40            \\
\cline{1-4}
\end{tabular*}
\end{table}

\newpage

\begin{table}[t]
\caption{Main components of the wave functions of the peaks around 12 MeV 
in $^{18,20}$O (visible in Fig.~2). 
Only the real parts of the amplitudes which are larger in absolute value than 0.1, are reported. $\pi$ and 
$\nu$ label the proton and neutron amplitudes, respectively.} 
\vspace{0.5cm}
\label{table2}
\begin{tabular*}{\textwidth}{|l|rr|}
\cline{1-3}
                            & $^{20}$O       & $^{22}$O    \\ 
\cline{1-3}
$\nu$,1p$_{1/2}$ 2s$_{1/2}$ &  0.82          & 0.83        \\
$\nu$,1d$_{5/2}$ 1f$_{7/2}$ &  0.31          & 0.35        \\
$\nu$,1d$_{5/2}$ 2p$_{3/2}$ & -0.32          &             \\
$\pi$,1p$_{1/2}$ 2s$_{1/2}$ & -0.18          &             \\
$\nu$,2s$_{1/2}$ 2p$_{3/2}$ &                & 0.30        \\
$\nu$,2s$_{1/2}$ 2p$_{1/2}$ &                & 0.10        \\
\cline{1-3}
\end{tabular*}
\end{table}

\newpage

\begin{table}[t]
\caption{Same as Table 1, in the case of the peak around 14.2 MeV in
$^{20}$O. }
\vspace{0.5cm}
\label{table3}
\begin{tabular*}{\textwidth}{|l|r|}
\cline{1-2}
                               & $^{20}$O       \\ 
\cline{1-2}
$\nu$,1d$_{5/2}$ 1f$_{7/2}$     &   0.18        \\
$\nu$,2s$_{1/2}$ 2p$_{3/2}$     &  -0.63        \\
$\nu$,1p$_{3/2}$ 1d$_{5/2}$     &  -0.18        \\
\cline{1-2}
\end{tabular*}
\end{table}

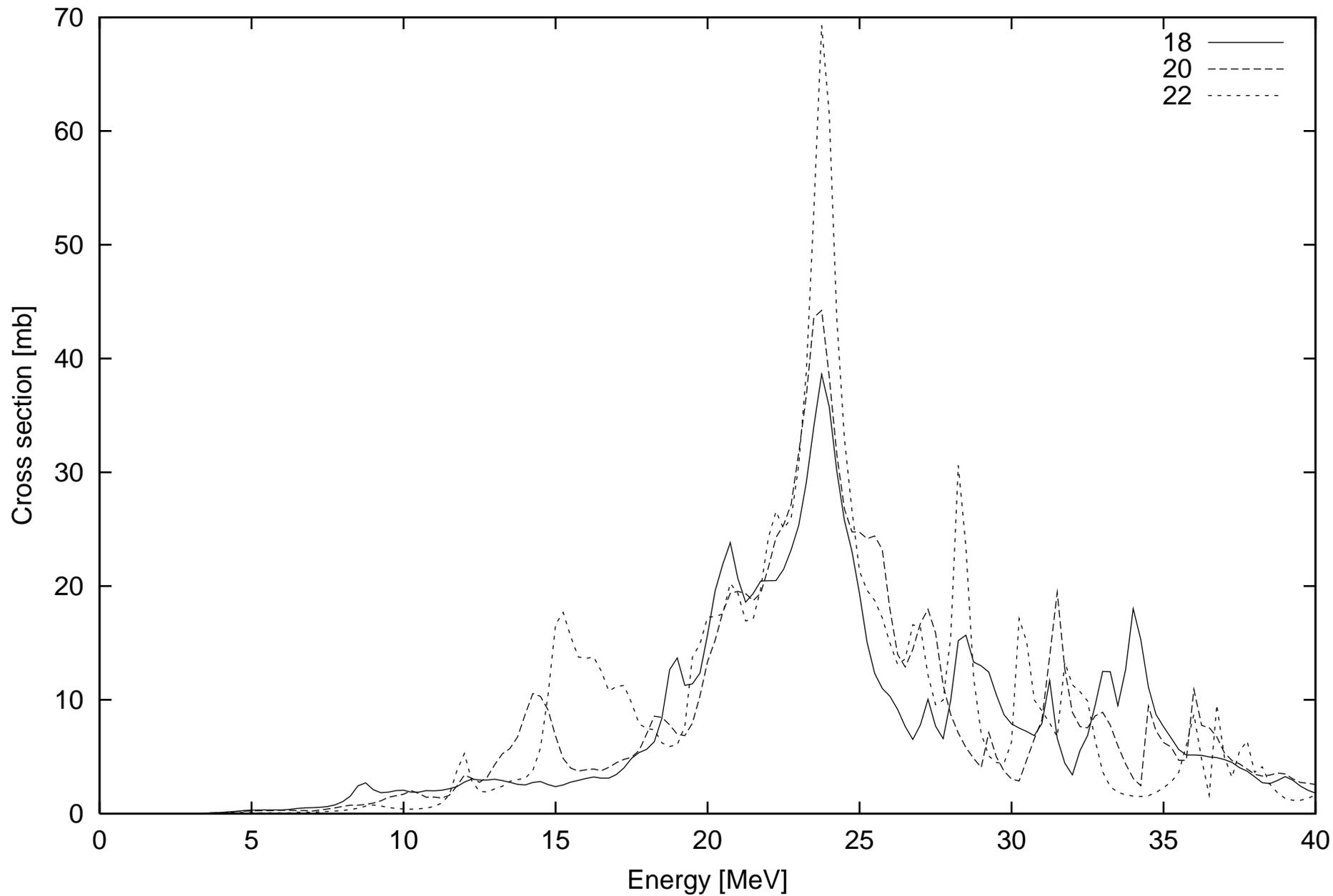
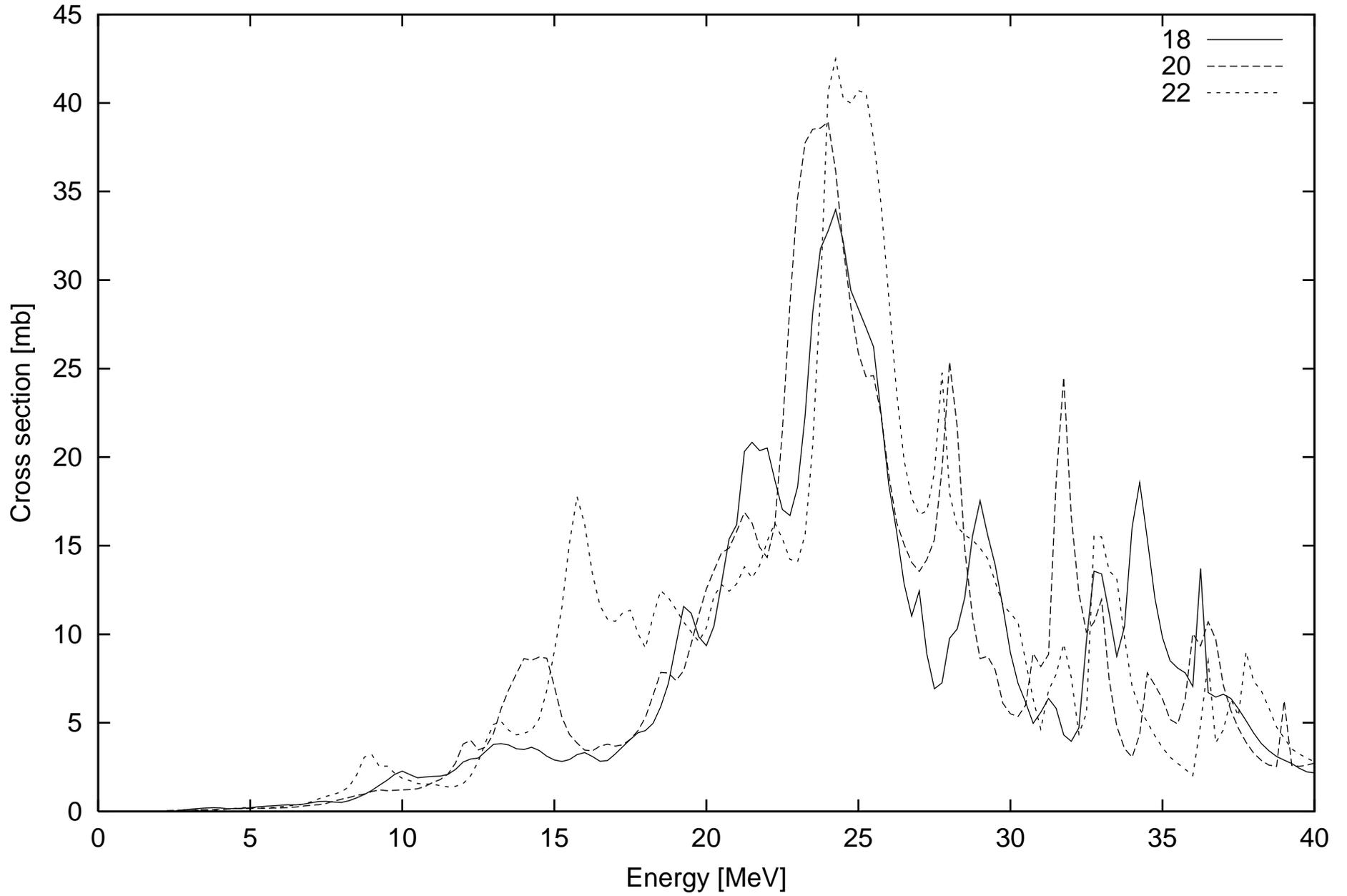
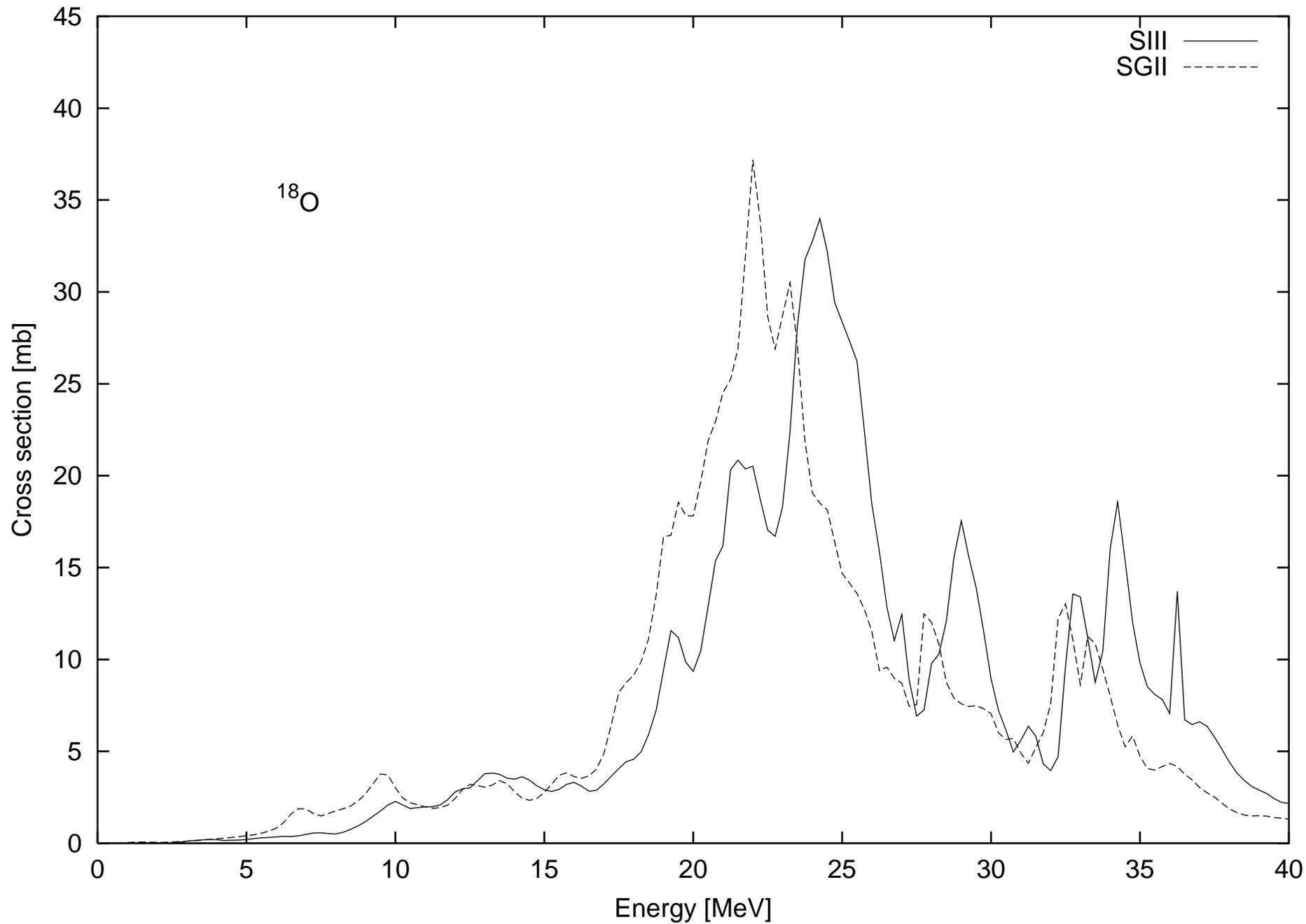
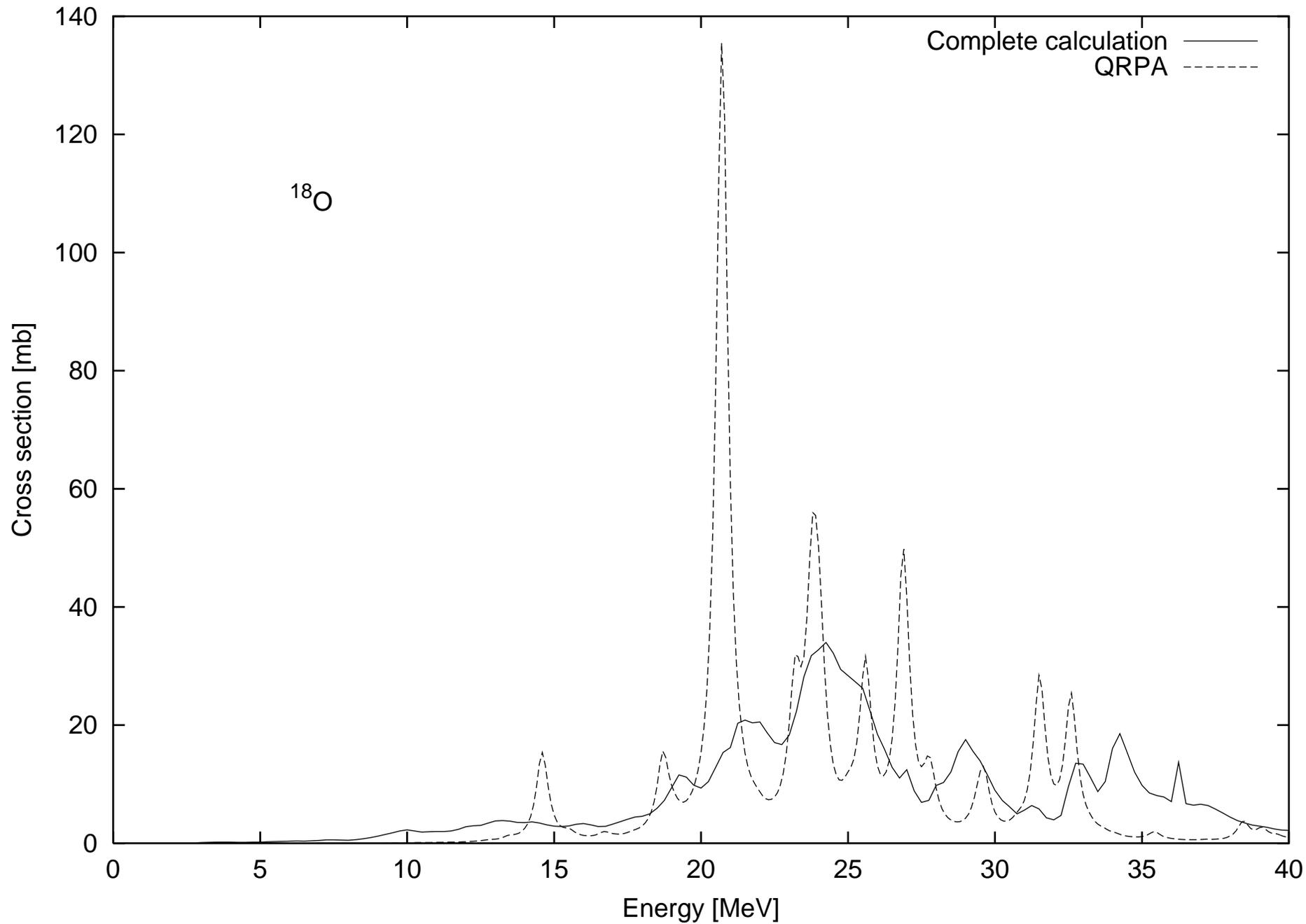
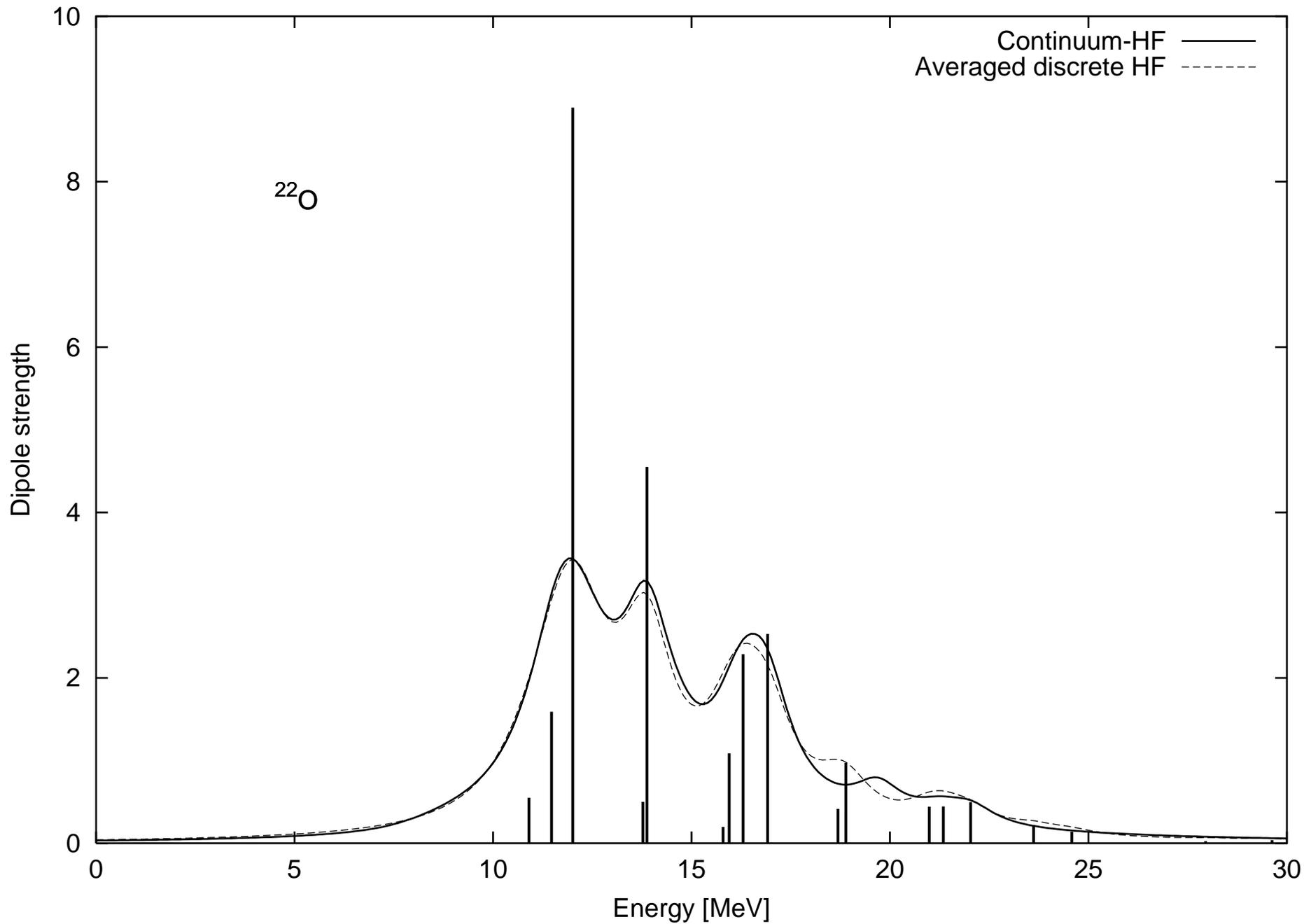
\end{document}